\documentstyle[11pt,aaspp4,psfig]{article}
\def\pp{\par\parshape 2 0truecm 15.5truecm 1truecm 14.5truecm\noindent}

\newcommand{\gtsima}{$\; \buildrel > \over \sim \;$}
\newcommand{\ltsima}{$\; \buildrel < \over \sim \;$}
\newcommand{\simgt}{\lower.5ex\hbox{\gtsima}}
\newcommand{\simlt}{\lower.5ex\hbox{\ltsima}}
\newcommand{\himpc}{{\hbox {$h^{-1}$}{\rm Mpc}} }
\newcommand{\zm}{z_{\rm max}}
\begin{document}
\begin{minipage}[c]{3cm}
  \psfig{file=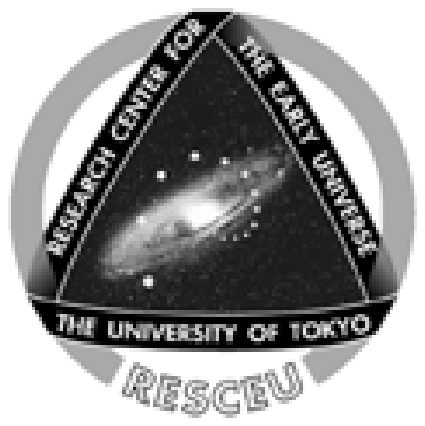,height=3cm}
\end{minipage}
\begin{minipage}[c]{9cm}
\begin{centering}
{
\vskip 0.1in
{\large \sf 
THE UNIVERSITY OF TOKYO\\
\vskip 0.1in
Research Center for the Early Universe}\\
}
\end{centering}
\end{minipage}
\begin{minipage}[c]{3cm}
\vspace{1.5cm}
RESCEU-32/97\\
UTAP-267/97
\end{minipage}\\

\title{Light-cone effect on higher-order clustering in redshift surveys}

\author{Takahiko Matsubara$^{1,2}$,  Yasushi Suto$^{1,2}$, and
Istv\'an Szapudi$^{3}$} 

\bigskip

\baselineskip13pt
\parskip5pt

\affil{\altaffilmark{1} Department of Physics, University of Tokyo,
  Tokyo 113, Japan} 
\affil{\altaffilmark{2} Research Center for the Early Universe
  (RESCEU), School of Science \\
  University of Tokyo, Tokyo 113, Japan} 
\affil{\altaffilmark{3} NASA/Fermilab Astrophysics Center, Fermi
  National Accelerator Laboratory \\ Batavia IL 60510-0500}

\medskip

\affil{\footnotesize e-mail: matsu@phys.s.u-tokyo.ac.jp, 
suto@phys.s.u-tokyo.ac.jp, szapudi@traviata.fnal.gov}

\bigskip

\centerline{Received August 12, 1997; accepted September 26, 1997}

\begin{abstract}
  We have evaluated a systematic effect on counts-in-cells analysis of
  deep, wide-field galaxy catalogues induced by the evolution of
  clustering within the survey volume. A multiplicative correction
  factor is explicitly presented, which can be applied after the
  higher order correlation functions have been extracted in the usual
  way, without taking into account the evolution.  The general theory
  of this effect combined with the ansatz describing the non-linear
  evolution of clustering in simulations enables us to estimate the
  magnitude of the correction factor in different cosmologies. In a
  series of numerical calculations assuming an array of cold dark
  matter models, it is found that, as long as galaxies are unbiased
  tracers of underlying density field, the effect is relatively small
  ($ \simeq 10\%$) for the shallow surveys ($z <0.2$), while it
  becomes significant (order of unity) in deep surveys ($z
  \sim1$). Depending on the scales of interest, the required
  correction is comparable to or smaller than the expected errors of
  on-going wide-field galaxy surveys such as the SDSS and 2dF.
  Therefore at present, the effect has to be taken into account for
  high precision measurements at very small scales only, while in
  future deep surveys it amounts to a significant correction.
\end{abstract}

\keywords {cosmology: theory - dark matter - large-scale structure of
universe -- galaxies: distances and redshifts }

\centerline{\sl The Astrophysical Journal (Letters), in press}

\baselineskip15pt
\section{Introduction}

Cosmological observations are necessarily carried out on a null
hypersurface or a light-cone. At low redshifts ($z<0.1$), this can be
regarded as to provide information on the constant-time hypersurface
($z=0$) which is a quite conventional implicit approximation
underlying cosmological studies using the galaxy redshift surveys.
When the depth of the survey volume exceeds $z\sim 0.1$, however, this
approximation breaks down, and one should simultaneously take account
of the intrinsic evolution of galaxy clustering and the light-cone
effect in addition to any other selection effect in interpreting the
data.  This is indeed the case for the on-going wide-field surveys of
galaxies including 2dF (2-degree Field Survey) and SDSS (Sloan Digital
Sky Survey).

To our knowledge, the first quantitative consideration of the
light-cone effect is made by Nakamura, Matsubara, \& Suto (1998) who
derived the systematic bias in the estimate of $\beta \approx
\Omega_0^{0.6}/b$ from magnitude-limited surveys of galaxies combining
the cosmological redshift distortion effect (Matsubara \& Suto 1996)
and the evolution of galaxy clustering within the survey volume.  In
this paper, we examine the light-cone effect on higher-order
statistics of galaxy clustering, considering counts-in-cells analysis
specifically.

Let us consider first the higher-order statistics on the idealistic
constant-time hypersurface. Denote the volume averaged $N$-th order
correlation functions at a redshift $z$ by $\overline{\xi}_N(R;z)$,
where $R$ is the comoving smoothing length, and introduce the
normalized higher-order moments $S_N(R;z) \equiv
\overline{\xi}_N(R;z)/[\overline{\xi}_2(R;z)]^{N-1}$.  The
hierarchical clustering ansatz states that $S_N(R;z)$ is constant and
independent of the scale $R$. This is a good approximation in
nonlinear regimes, although small but definite scale-dependence is
clearly detected from N-body experiments (Lahav et al. 1993; Suto
1993; Matsubara \& Suto 1994; Suto \& Matsubara 1994; Jing \&
B\"{o}rner 1997). In addition, perturbation theory predicts that
$\overline{\xi}_N(R;z)$ evolves in proportion to
$\left[\overline{\xi}_2(R;z)\right]^{N-1}$, and therefore $S_N(R;z)$
is independent of time, i.e. it is constant with respect to $z$.

The next section describes the general theory of the light cone effect
on $S_N(R;z)$ defined above.  Using the ansatz by Jain, Mo, \& White
(1995; hereafter JMW), \S3 evaluates the appropriate correction in an
array of cold dark matter (CDM) models. Finally, \S4 summarizes the
results and discusses the implications for redshift surveys.

\section{Observing the higher-order moments on the light-cone}

It is difficult to estimate $\overline{\xi}_N(R;z)$ observationally
since $z$ is changing over the volume of galaxy sample. While in
principle one could measure the $N$-point functions on $z = const$
surfaces, in practice this would result in a diminished volume, thus a
significant increase of the errors.  Instead it is more practical to
extract the following $N$-th order correlation functions averaged over
the volumes on the light-cone:
\begin{equation}
  \overline{\xi}_N(R;<\zm) \equiv {\displaystyle 
\int_0^{\zm} z^2 dz ~w(z) \overline{\xi}_N(R;z) \,  \over
\displaystyle \int_0^{\zm} z^2 dz ~w(z)} .
  \label{eq:xinzm}
\end{equation}
In the above expression, we assume that the observation is performed
with the fixed solid angle, and the sampling cells for the analysis
are placed randomly in $z$-coordinate with $w(z)$ being their
weighting function. If the cells were located randomly in the comoving
coordinates, the volume element $z^2dz$ should have been replaced by
$d_A(z;\Omega_0,\lambda_0)^2 c|dt/dz|dz$ ($d_A$ is the angular
diameter distance; see, Nakamura et al. 1998) and thus the procedure
itself becomes dependent on adopted values of $\Omega_0$ and
$\lambda_0$.  In principle $w(z)$ is an arbitrary function, and should
be determined so as to maximize the signal-to-noise ratio given the
selection function of individual observation. By $\zm$ we denote the
redshift corresponding to the depth of the survey. For a
volume-limited sample, for instance, it is natural to set $w(z)=1$ and
$\zm$ as the maximum redshift of the sample.  Similarly we define the
(observable) higher-order moments averaged over the light-cone as
\begin{equation}
 \overline{S_N}(R;<\zm) \equiv
{\overline{\xi}_N(R;<\zm)
\over \left[\overline{\xi}_2(R;<\zm)\right]^{N-1}} .
  \label{eq:snzm}
\end{equation}

It is useful to introduce the function $G(z)$ which describes the
evolution of the averaged two-point correlation function:
\begin{equation}
  \overline{\xi}(R;z) = G(z) \overline{\xi}(R;0) .
  \label{eq:gz}
\end{equation}
In linear regime, $G(z)$ is equivalent to $[D(z)/D(0)]^2$ where
$D(z)=D(z;\Omega_0,\lambda_0)$ is the linear growth rate.  Although
the above relation (\ref{eq:gz}) does not exactly hold in the
nonlinear regime, several approximation formulae are derived in the
literature, which empirically describe the evolution by allowing
$G(z)$ depend on the scale $R$ (see \S 3 for details).

Once we accept the evolution law (\ref{eq:gz}), equation
(\ref{eq:snzm}) is explicitly written as
\begin{equation}
 \overline{S_N}(R;<\zm)  = 
  \frac{\displaystyle
     \int_0^{\zm} z^2 dz~w(z) S_N(R;z) \left\{G(z)\right\}^{N-1}
      \left[\int_0^{\zm} z^2 dz ~w(z)\right]^{N-2}}
 {\displaystyle \left[\int_0^{\zm} z^2 dz ~w(z) G(z)\right]^{N-1}} .
\label{eq:m4}
\end{equation}
If $\zm \ll 1$, the above expression is expanded in terms of $\zm$ as
follows:
\begin{eqnarray}
&&
 \overline{S_N}(R;<\zm)  = 
  S_N(R;0) + \frac34 S_N'(0) \zm
\nonumber\\
&&
\qquad +\;
  \left[
    \frac3{160} (N-1)(N-2) S_N(0) G'(0)^2 + 
    \frac3{80}(N-1) S_N'(0) G'(0) +
    \frac{3}{10} S_N''(0)
  \right] \zm^2
\nonumber\\
&&
\qquad +\;
 {\cal O}(\zm^3) ,
\label{eq:m5}
\end{eqnarray}
where $S_N'(0)$ denotes $\partial S_N(R;z)/\partial z|_{z=0}$ and so
on. The above expansion up to ${\cal O}(\zm^2)$ is valid as long as
the weighting function is well-approximated up to the same order:
\begin{equation}
w(\zm) = w(0) + w'(0)\zm + {1\over 2} w''(0)\zm^2 .
\end{equation}
In other words, $\zm$ should be set to be smaller than the effective
window size of $w(\zm)$.

It is interesting to note that up to ${\cal O}(\zm^2)$ equation
(\ref{eq:m5}) is independent of $w(z)$, and that ${\cal O}(\zm)$ term
is determined only by $S_N'(0)$ independently of $G(z)$.  Since
$S_N'(0)$ is expected to vanish in linear theory (Fry 1984; Goroff et
al.  1986; Bernardeau 1992), and shown to be relatively small even in
quasi- and fully- nonlinear regimes (Bouchet et al. 1992; Lahav et al.
1993; Colombi, Bouchet, \& Hernquist 1995; Szapudi et al.  1997),
equation (\ref{eq:m5}) implies that the light-cone effect is very
small for 2dF and SDSS galaxy redshift surveys ($\zm <0.2$). It should
be noted, however, that if galaxies are biased relative to the mass
density field, $S_N'(0)$ may not be necessarily small.  So any signal
proportional to $\zm$ provides a clear indication of the
time-dependent biasing of galaxies (see e.g., Fry 1996; Mo \& White
1996; Mo, Jing \& White 1997).

\section{Evaluating the light-cone effect: an example}

In order to evaluate the effect of observational average on the
light-cone, we assume that $S_N(R;z)$ does not evolve with $z$, i.e.,
$S_N(R;z) = S_N(R;0)$. As described above, this is a reasonable
approximation as long as galaxies are unbiased tracers of underlying
density field.  If we introduce the measure of the light-cone effect:
\begin{equation}
  \Delta_N(R;<\zm) \equiv {\overline{S_N}(R;<\zm) \over S_N(R;0)} - 1 ,
  \label{eq:deltanz}
\end{equation}
equations (\ref{eq:m4}) and (\ref{eq:m5}) with $S_N(R;z)=S_N(R;0)$ reduce
to
\begin{equation}
  \Delta_N(\zm) =
    \frac3{160} (N-1)(N-2) G'(0)^2 \zm^2 + {\cal O}(\zm^3) .
  \label{eq:m6}
\end{equation}
Note that $(1+\Delta_N)$ can be regarded as a correction factor as
well, if one measures the $S_N$'s {\em without} considering the
evolution of clustering.  This constitutes a simple and practical
method, which we propose for future measurements, when compensation
for the light cone effect is required.

To evaluate equation (\ref{eq:m4}), we need a model for $G(z)$. For
this purpose, we adopt the ansatz originally put forward by Hamilton
et al. (1991) and improved later by Peacock \& Dodds (1994, 1996) and
JMW. To be specific, we apply the fitting formula by JMW which relates
the evolved two-point correlation function $\overline{\xi}_E(R;z)$
with its linear counterpart $\overline{\xi}_L(R_0;z)$ as follows:
\begin{eqnarray}
  \label{eq:jmwfit1}
\overline{\xi}_E(R;z) &=& B(n) \, F[\overline{\xi}_L(R_0;z)/B(n)] ,\\
  \label{eq:jmwfit4}
 F(x) &=& {x+0.45 x^2 -0.02 x^5+0.05 x^6 \over 1+0.02 x^3 + 0.003 x^{9/2}} .
\end{eqnarray}
In the above equations, $n$ denotes a power-law index of the power
spectrum, $R_0 =[1 + \overline{\xi}_E(z,R)]^{1/3} R$, and $B(n) =
[(3+n)/3]^{0.8}$. JMW show that the above formula works reasonably
well even for CDM models by replacing $n$ by the effective spectral
index evaluated at the scale which is just entering nonlinear
regime. In general the resulting $n$ depends on $z$, which we neglect
below for simplicity; for galaxy surveys which we are primarily
interested in, the $z$-dependence of $n$ near $z=0$ is expected to be
very small.

The inverse of equation (\ref{eq:jmwfit1}) is formally written as
$\overline{\xi}_L(R_0;z) = B(n) F^{-1}[\overline{\xi}_E(R;z)/B(n)]$
and JMW's empirical fit to $F^{-1}(y)$ is
\begin{eqnarray}
F^{-1}(y)= y\left( {1+ 0.036 y^{1.93} +0.0001 y^3
       \over   1+1.75 y -0.0015 y^{3.63} +0.028 y^4}
      \right)^{1/3} .
  \label{eq:jmwinv2}
\end{eqnarray}
Then $\overline{\xi}_E(R;z)$ is expressed explicitly in terms of
$\overline{\xi}_E(R;0)$:
\begin{equation}
  \overline{\xi}_E(R;z) = 
  B(n)   F\left[ {D^2(z) \over D^2(0)} 
          F^{-1}[\overline{\xi}_E(R,0)/B(n)]
          \right] .
  \label{eq:m12}
\end{equation}

Let us introduce a parameter $\alpha(R) \equiv
F^{-1}\left[\overline{\xi}_E(R,0)/B(n)\right]$ which characterizes the
variance on a scale $R$ at $z = 0$, and thus depends on $\Omega_0$ and
$\lambda_0$ through the shape of the fluctuation spectrum.  Then the
scale-dependent evolution factor $G(z)=G(R;z)$ in equation
(\ref{eq:gz}) is given by
\begin{equation}
G(R;z) \equiv  
{\overline{\xi}_E(R;z) \over \overline{\xi}_E(R;0)}
= \frac{1}{F(\alpha)} F\left[{D^2(z) \over D^2(0)}  
                         \alpha \right] .
  \label{eq:grz}
\end{equation}
For the convenience of $z$-expansion, we calculate the derivatives of
the above quantity at $z = 0$:
\begin{eqnarray}
  \left.{\partial G(R;z) \over \partial z}\right|_{z=0}
    &=& - 2 f_0 \frac{\alpha F'(\alpha)}{F(\alpha)},
  \label{eq:m14}
\\
  \left.{\partial^2 G(R;z) \over \partial z^2}\right|_{z=0}
      &=& 4 f_0^{\,2} \frac{\alpha^2 F''(\alpha)}{F(\alpha)} +
  \left(2 f_0^{\,2} + 2 f_0 q_0 + 3 \Omega_0 \right)
  \frac{\alpha F'(\alpha)}{F(\alpha)},
\label{eq:m15}
\end{eqnarray}
where $f_0 = d\ln D/da|_{z=0}$, $q_0 = \Omega_0/2 - \lambda_0$.  The
above expressions indicate how the light-cone effect depends on
$\Omega_0$ and $\lambda_0$ at $\zm \ll 1$.  Note, that they are
involved in ${\cal O}(\zm^2)$ term and thus do not contribute
significantly at small $z$.

\section{Results and conclusions}

\begin{figure}
\begin{center}
    \leavevmode\psfig{file=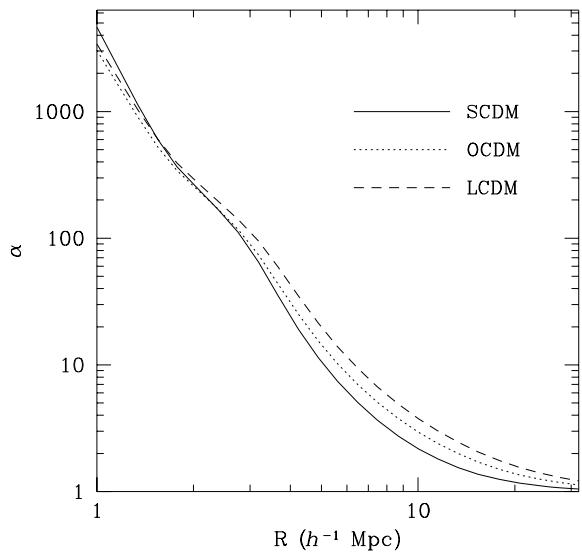,width=10cm}
\end{center}
\caption{ $\alpha(R)$ is plotted against $\log_{10}R(1\himpc)$ 
  for SCDM (solid line), OCDM (thin dotted line), and LCDM (thick
  dotted line) summarized in Table 1.
\label{fig:ar}}
\end{figure}

\begin{figure}
\begin{center}
    \leavevmode\psfig{file=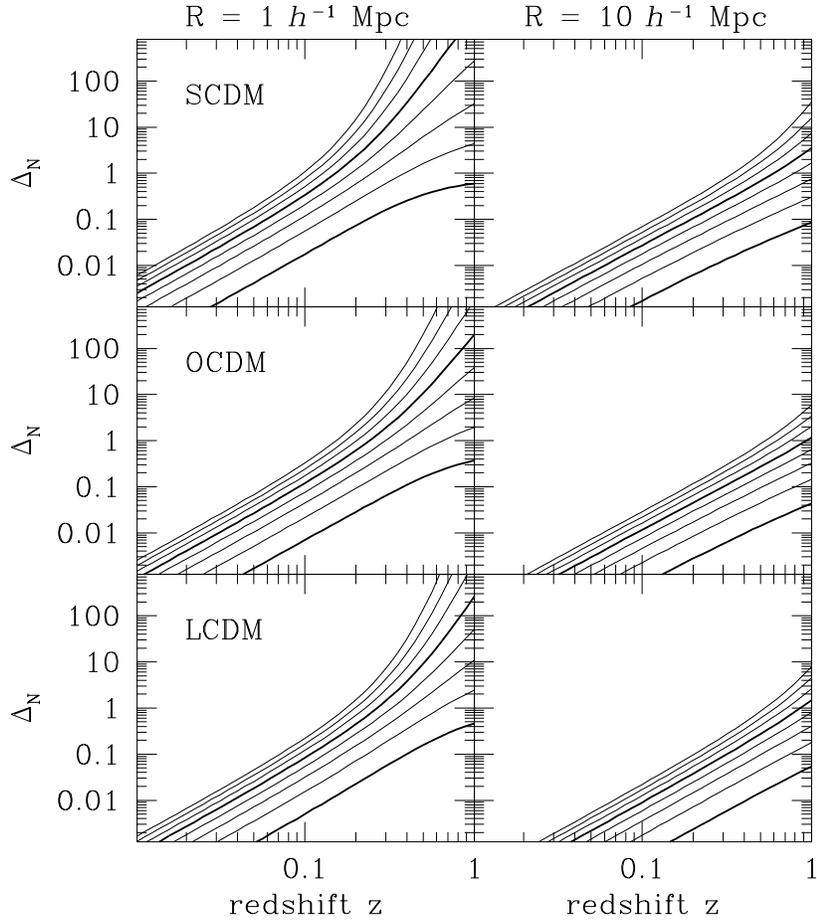,width=14cm}
\end{center}
\caption{$\log_{10}\Delta_N(R;z)$ are shown as functions of $\log_{10}z$ at
  $R=1\himpc$ ({\it left panels}) and $10\himpc$ ({\it right panels});
  SCDM ({\it top panels}), OCDM ({\it middle panels}), and LCDM ({\it
    bottom panels}). The family of curves display different
   orders from $N = 3\ldots N = 10$ monotonically
  upward; for $N = 3$, and $N = 7$ is plotted with thick lines
  for orientation. 
\label{fig:dnz}}
\end{figure}

\begin{figure}
\begin{center}
    \leavevmode\psfig{file=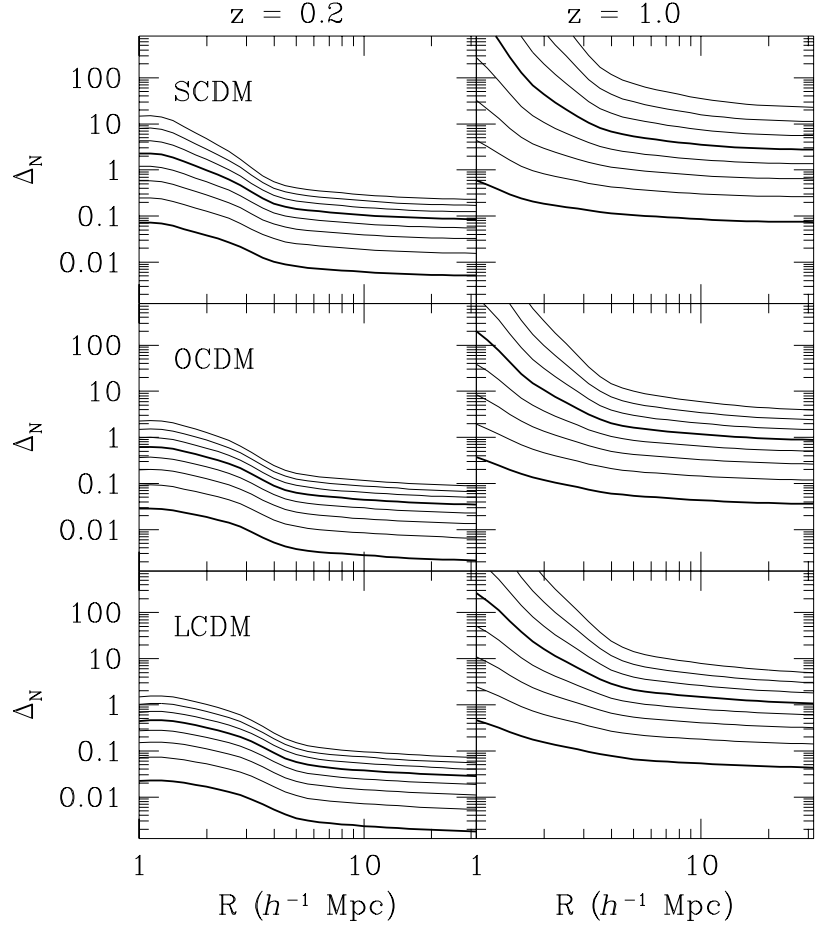,width=14cm}
\end{center}
\caption{$\log_{10}\Delta_N(R;z)$ are displayed as functions 
  of $\log_{10}R$ at $z=0.2$ ({\it left panels}) and $1.0$ ({\it right
    panels}); SCDM ({\it top panels}), OCDM ({\it middle panels}), and
  LCDM ({\it bottom panels}). The family of curves is the same as for
  Fig.2.
\label{fig:dnr}}
\end{figure}

Using equations (\ref{eq:m4}) and (\ref{eq:grz}) and assuming $S_N(z)=
S_N(0)$, we can evaluate the evolutionary effect on
$\overline{S_N}(R;<\zm)$ or $\Delta_N(<\zm)$. As examples, we consider
three representative CDM models (Table 1) whose fluctuation amplitude
$\sigma_8$ is normalized so as to reproduce the abundances of clusters
of galaxies (e.g., Kitayama \& Suto 1997; Kitayama, Sasaki \& Suto
1997). The results are displayed on a series of figures.  Figure
\ref{fig:ar} shows how $\alpha$ is related to the comoving smoothing
length $R$ in these models.  Figure \ref{fig:dnz} displays
$\Delta_N(R;z)$ as a function of $z$, and finally Figure
\ref{fig:dnr} plots $\Delta_N(R;z)$ against $R$.

The general appearance of the figures suggests that the light-cone
effect is a fairly robust feature, although its details depend on the
model. In all cases SCDM appears to give the strongest effect, while
for OCDM and LCDM it is slightly less pronounced. Nevertheless the
difference is fairly small and qualitatively all models behave
similarly.  Note also, that the magnitude of the correction depends on
the order $N$, and, in accord with intuition, it is monotonically
increasing for higher order. As expected, the light-cone effect
becomes larger as $\zm$ increases, which can be seen in Figure
\ref{fig:dnz}. Although the correction is relatively small for shallow
surveys with $z \simlt 0.2$ samples, $\Delta_N(R;<\zm)$ becomes
$\simgt 10$\% in nonlinear scales ($R\sim 1\himpc$). In SCDM, for
instance, $\Delta_N(R;<\zm)$ exceeds unity for $N \geq 6$ for the
entire dynamic range plotted.  Furthermore Figure \ref{fig:dnr}
indicates that even if the hierarchical ansatz is correct, i.e.,
$S_N(R;z)$ is independent of $R$, the light-cone effect should
generate apparent scale-dependence, since the correction behaves
differently at different scales at a given redshift.

The future SDSS will be able to measure the moments of the galaxy
density field with unprecedented accuracy. Unless unforeseen
systematics exists, it will determine them with less than a few
percent error for $N \leq 3$ and $10\%$ for $N = 4$ between $1$ and
$50\himpc$ (see Colombi, Szapudi, \& Szalay 1997 for
details). According to Figures \ref{fig:dnz} and \ref{fig:dnr}, the
light cone effect will be much smaller than these errors, or at most
of the same order, depending on the scales and models. The correction
could be potentially non-negligible only at the smallest scales. A
similar conclusion can probably be drawn about the 2dF survey.  On the
other hands, for future deep surveys, which should aim at smaller
scales especially if carried out by the Next-Generation Space
Telescope, our calculations will be of utmost importance.  According
to Figure \ref{fig:dnr} the required correction can range from up to
unity for $S_3$ through factors of few for $S_6$ to factors of hundred
for $S_{10}$.

\bigskip

This research was supported in part by the Grants-in-Aid for the
Center-of-Excellence (COE) Research of the Ministry of Education,
Science, Sports and Culture of Japan (07CE2002) to RESCEU (Research
Center for the Early Universe), University of Tokyo.
I.S. was supported by DOE and NASA through grant
NAG-5-2788 at Fermilab.

\bigskip
\clearpage

\baselineskip=13pt
\parskip4pt
\bigskip
\centerline{\bf REFERENCES}

\bigskip

\def\apjpap#1;#2;#3;#4; {\pp#1, {#2}, {#3}, #4}
\def\apjbook#1;#2;#3;#4; {\pp#1, {#2} (#3: #4)}
\def\apjppt#1;#2; {\pp#1, #2.}
\def\apjproc#1;#2;#3;#4;#5;#6; {\pp#1, {#2} #3, (#4: #5), #6}

\apjpap Bernardeau, F. 1992;ApJ;392;1;
\apjpap Bouchet, F.R., Juszkiewicz, R., Colombi, S. \& Pellat, R., 1992;
ApJ;394;L5;
\apjpap Colombi, S., Bouchet, F.R., \& Hernquist,L. 1995;ApJ;281;301; 
\apjppt Colombi, S., Szapudi, I. Szalay, A.S., 1997;MNRAS, submitted;
\apjpap Fry, J.N. 1984;ApJ;277;L5;
\apjpap Fry, J.N. 1996;ApJ;461;L65;
\apjpap Goroff,M.H., Grinstein,B., Rey, S.J., \& Wise, M.B. 1986;
   ApJ;311;6;
\apjpap Hamilton, A.J.S., Kumar, P., Lu, E., \& Matthews, A. 1991;
  ApJ;374;L1;
\apjpap Jain, B., Mo, H.J., \& White, S.D.M. 1995;MNRAS;276;L25 (JMW);
\apjpap Jing, Y.P. \& B\"{o}rner, G. 1997;A\&A;318;667;
\apjpap Kitayama, T., \& Suto, Y. 1997;ApJ;490;in press
(astro-ph/9702017);
\apjppt Kitayama, T., Sasaki, S., \& Suto, Y. 1998;
Pub.Astron.Soc.Japan, in press (astro-ph/9708088);
\apjpap Lahav, O., Itoh, M., Inagaki, S., \& Suto, Y. 1993;
  ApJ;402;387;
\apjpap Matsubara, T. \& Suto, Y. 1994;ApJ;420;497;
\apjpap Matsubara, T. \& Suto, Y. 1996;ApJ;470;L1 (astro-ph/9604142);
\apjpap Mo, H.J., \& White, S.D.M. 1996;MNRAS;282;347;
\apjpap Mo, H.J., Jing, Y.P., \& White, S.D.M. 1997;MNRAS;284;189;
\apjpap Nakamura, T.T., Matsubara, T. \& Suto, Y. 1998;
   ApJ;493;in press (astro-ph/97060349);
\apjpap Peacock, J.A. \& Dodds, S.J. 1994;MNRAS;267;1020;
\apjpap Peacock, J.A. \& Dodds, S.J. 1996;MNRAS;280;L19;
\apjbook Peebles, P. J. E. 1980; The Large Scale Structure of the Universe;
   Princeton University Press; Princeton;
\apjpap Suto, Y. 1993;Prog.Theor.Phys.;90;1173;
\apjpap Suto, Y. \& Matsubara, T. 1994;ApJ;420;504;
\apjppt Szapudi, I.,  Quinn, T., Stadel, J., \& Lake, G. 1997; 
in preparation;

\clearpage

\begin{table}
\caption{CDM model parameters}
\begin{center}
\begin{tabular}{crrcr}
\\
\tableline
Model & $\Omega_0$ & $\lambda_0$ & $h$ & $\sigma_8$ \\
\hline
\hline
SCDM & $1.0$ & $0.0$ & $0.5$ & 0.6 \\ 
OCDM & $0.45$ & $0.0$ & $0.7$ & $0.8$ \\ 
LCDM & $0.3$ & $0.7$ & $0.7$ & $1.0$ \\ 
\hline
\end{tabular} 
\end{center}
\end{table} 

\end{document}